# Performance Evaluation of a Deployed 4G LTE Network

E.T. Tchao[1]

Dept. of Computer Engineering, KNUST
Kwame Nkrumah Univ. Science and Tech.,
Kumasi, Ghana

J.D. Gadze[2], Jonathan Obeng Agyapong[3]

Dept. of Electrical/Electronic Engineering
Kwame Nkrumah University of Science and Tech.,
Kumasi, Ghana

*Abstract*—In Ghana and many countries within Sub-Sahara Africa, Long Term Evolution (LTE) is being considered for use within the sectors of Governance, Energy distribution and transmission, Transport, Education and Health. Subscribers and Governments within the region have high expectations for these new networks and want to leverage the promised enhanced coverage and high data rates for development. Recent performance evaluations of deployed WiMAX networks in Ghana showed promising performance of a wireless broadband technology in supporting the capacity demands needed in the peculiar Sub-Saharan African terrain. The deployed WiMAX networks, however could not achieve the optimal quality of service required for providing a seamless wireless connectivity demands needed for emerging mobile applications. This paper evaluates the performance of some selected key network parameters of a newly deployed LTE network in the 2600 MHz band operating in the peculiar Sub-Saharan African terrain under varied MIMO Antenna Configurations. We adopted simulation and field measurement to aid us in our evaluation. Genex Unet has been used to simulate network coverage and throughput performance of 2X2, 4X4 and 8X8 MIMO configurations of the deployed networks. The average simulated throughput per sector of 4X4 MIMO configuration was seen to be better than the 2X2 configuration. However, the percentage coverage for users under the 2x2 MIMO simulation scenario was better than that of the adaptive 4x4 MIMO configuration with 2x2 MIMO achieving 60.41% of coverage area having throughput values between 1 - 40Mbps as against 55.87% achieved by the 4x4 MIMO configuration in the peculiar deployment terrain.

## I. INTRODUCTION

Fourth Generation Long Term Evolution (4G-LTE) being a technology for data communication designed by 3rd Generation Partnership Project is meant to provide high speed and low latency mobile wireless connectivity over long distances. LTE has a flexible spectrum supporting the bands ranging from 450 MHz to 4.5 GHz. LTE also support bandwidth ranging from 1.25 MHz to 20MHz [1].

LTE has a flat Internet Protocol (IP) Based architecture. To able to achieve this, it is deployed with Orthogonal Frequency Division Multiple Access (OFDMA) in the downlink. The uplink is implemented using Single Carrier Division Multiple Access. This helps to enhances high data rate. Another feature in 4G LTE which enhances high throughput over long distances is Multiple Input Multiple Output (MIMO) systems. MIMO is a multi-antenna technology which uses $M$ multiple antennas at the transmitting and $N$ multiple antennas at the receiving ends. This configuration can be denoted by a $MxN$ MIMO system. Increasing the number of transmitting antennas increases the average signal to noise ratio per receiver which enhances cell range and throughput capacity in MIMO systems [2].

A simplified 4G LTE architecture is generally made up of the evolved Packet Core, the eNodeB and the User equipment. The user connects to the network via the wireless air interface to the eNodeB. This wireless medium is enhanced by MIMO which allows multiple data streams to be transmitted and received concurrently through multiple antennas [3].

The competitive advantage that LTE therefore has is its ability to provide a very high throughput and longer coverage for the end users [4]. When a user is connected to a 4G LTE network, then he is connected to a network with a flat IP-based architecture which uses OFDMA and SCDMA multiple access in the downlink and uplink respectively and employs MIMO in the wireless medium. This enables the technology to provide wireless broadband internet access to its end users even in remote areas.

## II. 4G LTE DRIVES DEMAND

There has been a significant increase in mobile subscription over the past five years. According to Ericsson's mobility report for June 2016 which provides statistics for mobile data subscription for the past five years and a forecast into the next five years, the total number of mobile subscription as at the first quarter of 2016 was up to 7.4 billion [5]. Their analysis showed that, mobile broadband subscription increases continuously every year whiles fixed broadband subscriptions remains nearly constant over the years. Gee Rittenhouse [6] also predicts a 1000X of data by the year 2020. His estimation showed an increasing demand for data services for the past five years and a greater demand is predicted for the next five years ahead.

4G LTE promises high wireless mobile data rate of up to 300Mbps depending on the MIMO type which far exceeds the throughput of the preceding Third Generation (3G) mobile wireless access technology [7]. 4G LTE subscriptions reached 1.2 billion in 2016 with about 150 million new subscriptions and an estimated subscription of up to 4.3 billion by the end of 2021.

In developed markets there has been a large migration to this new technology resulting in decline in the use of EDGE/GSM [8]. However, in developing market, such as that of Ghana and many countries within the Sub-Saharan Africa sub-region, migrating to LTE is an option due to high cost of devices and network deployment. However, because of the high capacity network LTE assures, some countries within Sub-Saharan African have started deploying LTE solutions to provide mobile wireless connectivity to serve eager customers who have ever increasing broadband internet demands. Ghanaians have only just begun to connect to these newly deployed 4G LTE networks to enjoy the quality of experience that the technology promises.

## III. RELATED WORKS

The network performance of LTE networks have been studied extensively in Literature. The authors in [9] presented a computation of path loss using different propagation models for LTE Advance networks in different terrains; rural, dense urban and suburban. The paper analyzed path loss for broadband channels at 2300MHz, 2600MHz and 3500 MHz using MATLAB. The paper performed this study using the following prediction models; Stanford University Interim model, COST 231 Walfisch-Ikegami model, ECC-33/Hata Okumura extended model and COST 231 Hata model. The comparison of the various propagation models showed that the least path loss was obtained by using COST-231 Hata prediction method. It was concluded that the prediction errors for SUI and ECC models are significantly more than COST 231 Hata and COST Walfisch-Ikegami models.

The research work published in [10] investigates three empirical propagation models for 4G LTE networks in the 2-3 GHz band. This was undertaken in urban and rural areas in Erbil city in Iraq. The results were compared with field measurements and tuning methods were suggested to fit the measured path loss results of the Standford University Interim (SUI) model and Okumura Hata, extended COST 231 Hata model. It was seen that the optimum model which was closer to the measured path loss data was the extended COST 231 Hata model. COST 231 Hata model had the mean error value lowered to a zero value, the mean standard deviation value lowered to 7.8dB and root mean square error being 7.85dB. Thus the COST 231 Hata model was the propagation prediction model predicted for the city.

The research work published in [11] makes a comparison for the diverse suggested propagation models to be implemented for 4G wireless at various antenna heights. The path loss for the various models; Stanford University Interim model, COST 231 Hata model and COST Walfisch-Ikegami model were computed in different environmental scenarios; Urban, suburban and rural areas. MATLAB simulation was used for the computation for frequencies in the 2300 MHz, 2600 MHz and 3500MHz band. It was concluded that path loss was least using COST 231 Hata model for all environmental scenarios and the different frequencies than the other models. The advantages of this approach were in its adaptability to dissimilar environments by infusing the appropriate correction factors for diverse environments. It was concluded that path loss was least in urban areas for 1900

MHz and 2100 MHz frequencies using SUI model. COST231 gave the highest path loss for 1900 MHz and Ericsson 9999 predicted the highest path loss for 2100 MHz band.

The research work published in [12] presents a computation of path loss using different propagation models for WiMAX in urban environments. The paper analyzed path loss and took field measurement at 3500 MHz using MATLAB. The paper performed this study using the following prediction models; COST 231 Hata model, Stanford University Interim model, Ericsson model, ECC-33/Hata Okumura extended model and free space path loss model. Simulations were performed at different receiver height at 3m, 6m and 10m. It was inferred that the best results were obtained from ECC-33 and SUI models with respect to least path loss in urban environments.

The effect of MIMO configurations on network deployments have also been widely studied in Literature. Authors in [13] presented a deterministic approach for simulating bit error rate (BER) of MIMO antenna configuration but this time in the presence of multiple interferers in the Sub-Saharan African terrain. The pattern of radiation was simulated for the WiMAX Antenna by the authors using Genex-Unet and NEC. To model BER performance, a Monte Carlo strategy was employed in the presence of 2x2 and 4x4 in the sight of ten intruding Base Stations (BS). In the presence of a number of interferers, poor estimates for average BER were observed for 2x2 MIMO deployments. 4x4 MIMO antenna configurations with efficiently suppressed side lobes were proposed for future WiMAX deployment for a better BER performance.

In [14], the network capacity performance of 2x2, 4x4, 8x8 and 12x12 systems was analyzed by the authors using MATLAB for real-time simulations. The authors simulated the channel capacity of the different MIMO schemes against probability error at different signal to noise ratio; 0dB, 10dB, 20dB, 40dB and 60dB. It was concluded that the maximum channel capacity is achieved at 60dB and 0dB for 12x12 and 2x2 MIMO configurations respectively.

There is little information on the performance of 4G networks in the sub-region. Authors in [15] measured 9.62Mbps downlink throughput for a deployed WiMAX network under realistic network conditions in the Sub-Saharan African terrain. This throughput performance was poor when compared with reported network performance of deployed LTE networks in Europe and Asia. Documented results of the performance of LTE under various deployment scenarios in Europe and Asia were also presented in [16] and [17] respectively. In [16], the maximum reported measured downlink throughput was 52Mbps whiles a maximum throughput of 32Mbps was measured in [17] as compared to the simulated throughput of 60Mbps and 35Mbps respectively. Despite the variations between the measured and simulated results, an impressive performance of the deployed networks was realized confirming the assurance LTE gives.

Deployment of Long Term Evolution networks have just begun in Ghana and operators are very optimistic LTE could help address the limitations of the WiMAX networks. There is therefore the need to evaluate the performance of recently deployed networks to determine if they can deliver on their promised performance. It can also be seen from the related work however that, there are variations between promised results and actual performance of deployed network. Some of the related research works have studied the maximum theoretical values in terms of throughput which cannot best represent the realistic capacity that end users enjoy. The related work on coverage estimation studied the Path loss models but failed to apply the models to a typical deployment scenario to provide achievable cell range for the various antenna configurations. There is therefore the need to provide a realistic data on throughput and coverage which can fairly represent consumer experience an LTE subscriber will enjoy.

## IV. NETWORKING DIMENSIONING PROCESS

There have been several attempts aimed at deploying 4G networks in Ghana. Earlier pilot deployment of LTE and WiMAX networks in Ghana suffered from high levels of co-channel and adjacent channel interference from co-located Third generation antenna systems [18]. This high level of interference led to low capacity networks with limited connectivity in earlier deployed networks. This section retraces the steps followed by network and radio planning engineers in planning the first successful LTE network in Ghana.

Radio network planning is a significant process in all wireless propagation. It enables a planner to estimate expected network performance by the use of planning tools to validate theoretical models, technological requirements and business models. During the radio network planning stage of the network in the pilot area, the main goal was to simulate the throughput and the coverage that will best serve end users using theoretical models with the help of a radio planning tool. The dimensioning process which was used in successfully deploying the LTE under study involved a sequence of steps which served different requirements such as antenna radiation pattern simulation, coverage and capacity estimations to derive the final radio network plan for the 2X2 MIMO system. The dimensioning process is shown in Fig. 1. The dimensioning processes included:

- Selection of site. This should be closer to our target consumers and also have easy access to road and power source. Also, the type of area should be noted, whether urban, sub-urban or rural.

- Consideration of environmental factors: This involved analyzing the peculiar terrain which included vegetation, climate, height of buildings around the site, the topology of the land. This helps in considering how to tilt our antenna electrically and mechanically and the height of the transmitting antenna to be used.

- Frequency and interference planning: PUSC 1x3x3 is used for the deployment. There is therefore some probability of interference at high transmit powers so we simulated the added sidelobe suppression factor needed.

- Coverage Prediction analysis: This analysis is carried out in an outdoor environment to determine signal

strength within the coverage area and the network accessibility.

- Capacity Analysis: This enables prediction of the average throughput within the cell. It will also consider throughput distribution within the cells.

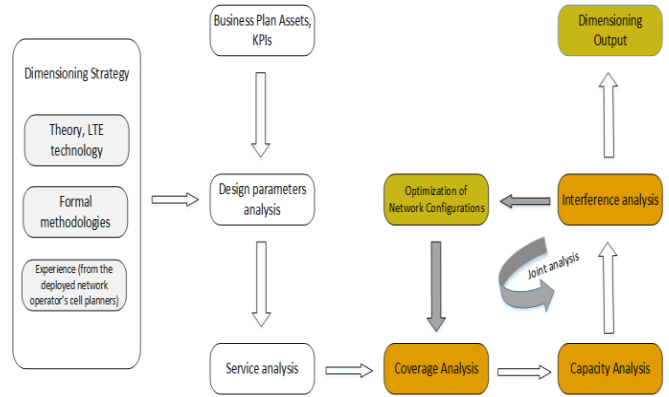

Fig. 1. Radio planning methodology.

The input parameters used for planning the network consisted of the business plan, the assets and the key performance indicators (KPIs). The first step of the dimensioning process obtained all the necessary information on the configurations which were used in the process. During the input analysis stage, theoretical and technological expertise was drawn from the network planning engineers who worked on deploying the initial WiMAX networks. This expert advice was adapted, taking into consideration of LTE technological standards, to implement the radio network plan for the LTE network.

The radio network planning was split into three main analysis namely; coverage, capacity and interference analysis. As part of the coverage analysis, the provided service areas were identified and the required number of LTE sites estimated. Further to the coverage analysis, the service areas had to be dimensioned to offer sufficient capacity as dictated by the number of users and services profiles. Therefore the next step was to estimate the required number of sectors to achieve this capacity. The estimates of LTE sites and sectors were then used to determine the configuration of the BS in the final joint analysis step. After the capacity and coverage analysis, the results of the initial configurations were evaluated followed by customizations to further improve the solution.

## V. SYSTEMS MODEL

This section presents and discusses the coverage and capacity models adopted to successfully dimension the network.

### A. Coverage Model

In this analysis, an N element MIMO antenna system in a one tier multicarrier system using a Partial Usage Sub-Channelization (PUSC), will be discussed. The frequency reuse scheme will be denoted by $N_c x N_s x N_f$, where:

- Nc is number of individual frequency channels in the LTE network.

- Ns is the number of sectors per cell.

- Nf is the number of segments in exploited frequency channel.

The deployed network uses a reuse scheme of PUSC 1x3x3. The layout of this frequency reuse scheme is shown in Fig. 2. The scheme uses three sector antennas and requires only one radio frequency (RF) channel for all sectors. The scheme uses three different sets of tones. Each of these tones is for a sector of a Base Station. This reduces significantly inter-cell interference and the outage area of the cell. Radio frequency (RF) planning is greatly made simple because the segments just have to be assigned to sectors while using the same RF channel among all Base Stations (BS).

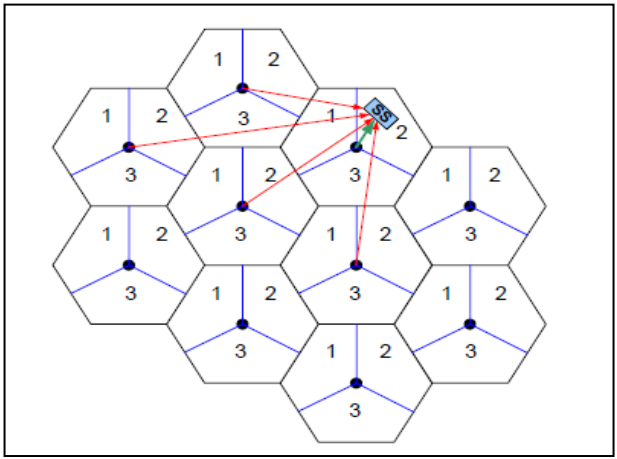

Fig. 2. PUSc 1x3x3 frequency reuse scheme.

Co-channel interference typically affects system configuration performance in the cell border areas for PUSC schemes. Authors in [19] evaluated the capacity of WiMAX MIMO systems in the cell borders for PUSC schemes and realized that, the degradation in network performance at cell borders were due to the presence of recorded high co-channel interference values. Results discussed in [20] showed that in a multicarrier system, each BS antenna can receive unsuppressed sidelobe emission from the antenna of adjacent cell BS. There was therefore the need to model the maximum range of the LTE network under evaluation by considering the effect of the unsuppressed sidelobe power emission that could come from neighboring cell sites during the radio planning stage.

In determining the maximum cell range, Friis transmission model played an essential role in the analysis and design of the network because when transmitting and receiving antennas are sufficiently separated by a considerable distance, Friis model would seek to make a relation between the power that is supplied to a transmitting antenna and the power that is received on the other end by a receiving antenna. The adopted transmission model is shown in Fig. 3. This model assumes that a transmitting antenna produces power density Wt($\theta_t$ , $\varphi_t$) in the direction ($\theta_t$ , $\varphi_t$).

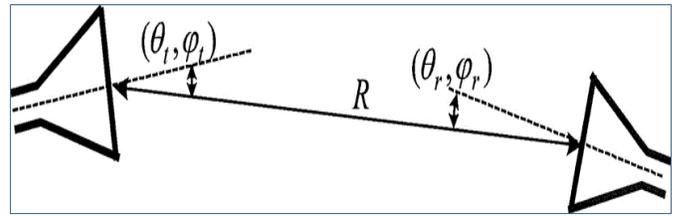

Fig. 3. Adopted wireless propagation model.

The transmitting antenna gain determines the power density in the given direction $G(\theta_t$ , $\varphi_t)$, on the distance $R$ between the point of observation and the antenna and the on the power that is supplied to the transmitter $P_t$. The power density can be modelled as [21]:

$$W_t = \frac{P_t}{4\pi R^2} G_t(\theta_t , \varphi_t) = \frac{P_t}{4\pi R^2} e_t D_t(\theta_t , \varphi_t) \qquad (1)$$

Here, et denotes the radiation efficiency of the transmitting antenna and Dt is its directivity. The receiving antenna has the power Pr at its terminals expressed by its effective area Ar and Wt:

$$P_r = A_r W_t \qquad (2)$$

To include polarization and heat losses in the receiving antenna, we add the radiation efficiency of the receiving antenna $e_r$ and the polarization loss factor (PLF) to the receive power equation as:

$$P_r = e_r \cdot PLF \cdot A_r W_t = A_r W_t \, e_r |\widehat{\mathbf{\rho}}_t \cdot \widehat{\mathbf{\rho}}_r|^2 \qquad (3)$$

$$\Rightarrow P_r = \underbrace{D_r(\theta_r , \varphi_r) \cdot \frac{\lambda^2}{4\pi}}_{A_r} \cdot e_r W_t |\widehat{\mathbf{\rho}}_t \cdot \widehat{\mathbf{\rho}}_r|^2 \qquad (4)$$

Here, Dr represents the directivity of the receiving antenna. When calculating PLF, the polarization vectors of the receiving and transmitting antennas, $\widehat{\mathbf{\rho}}_r \ and \ \widehat{\mathbf{\rho}}_t$, are evaluated based on their respective coordinate systems and so one of them has to be conjugated.

The signal is incident upon the receiving antenna from a direction ($\theta$r ,$\varphi$r), which is defined in the coordinate system of the receiving antenna. The received power can hence be modelled as:

$$P_r = D_r(\theta_r , \varphi_r) \cdot \frac{\lambda^2}{4\pi} \cdot \underbrace{\frac{P_t}{4\pi R^2} e_t D_t(\theta_t , \varphi_t)}_{W_t} \cdot e_r |\widehat{\mathbf{\rho}}_t \cdot \widehat{\mathbf{\rho}}_r|^2 \qquad (5)$$

The ratio of the received to the transmitted power is given by:

$$\frac{P_r}{P_t} = e_t e_r |\widehat{\mathbf{\rho}}_t \cdot \widehat{\mathbf{\rho}}_r|^2 \left(\frac{\lambda}{4\pi R}\right)^2 D_t(\theta_t , \varphi_t) D_r(\theta_r , \varphi_r) \qquad (6)$$

Once the impedance-mismatch loss factor is included in the transmitting and the receiving antenna systems, then the ratio above becomes:

$$\frac{P_r}{P_t} = (1 - |\Gamma_t|^2)(1 - |\Gamma_r|^2) e_t e_r |\widehat{\mathbf{\rho}}_t$$
$$\cdot \widehat{\mathbf{\rho}}_r|^2 \left(\frac{\lambda}{4\pi R}\right)^2 D_t(\theta_t , \varphi_t) D_r(\theta_r , \varphi_r) \qquad (7)$$

Where,

$\Gamma_t$ *and* $\Gamma_r$ are the reflection coefficients of the transmitting and receiving antenna,s respectively.

The efficient design of wireless systems and the approximation of antenna radiation efficiency when the antenna gain is established can be done using (6.7) which is a variation of Friis' transmission model.

In order to estimate the maximum cell detectable range, it is important to know all the specifications of the receiving and transmitting antenna like the gain, impedance-mismatch, polarization, losses and the minimum power at that the receiver would need to accurately operate $P_{r\,min}$ as well as the nominal power of the transmitter $P_t$ .The maximum range can then be calculated as:

$$R^2_{max} = (1 - |\Gamma_t|^2)(1 - |\Gamma_r|^2)e_t e_r |\hat{\rho}_t$$
$$\cdot \hat{\rho}_r|^2 \left(\frac{\lambda}{4\pi}\right)^2 \left(\frac{P_t}{P_{r\,min}}\right) D_t(\theta_t, \varphi_t) D_r(\theta_r, \varphi_r) \quad (8)$$

The receiver can reliably and effectively operate with a minimum power which is dependent on numerous factors, of which has Signal-to-noise-ratio (SNR) being the core one.

### B. Antenna Beamwidth and Sidelobe Power Modelling

The emergence of technology has greatly helped in antenna design because of the application of analytical and numerical optimization techniques. A number works related to the area of antenna array analysis and design have been presented in [22] and [23] where the optimization of the relative position of the antenna elements has been made possible by particle swarm optimization (PSO) as well as other well-known algorithms to get the minimum Sidelobe Levels (SLL) suppression factors. This part of the research summarizes the interference analysis stage which seeks to investigate the effect of antenna sidelobe emissions on multicarrier MIMO systems when the antenna beamwidth and side lobe level are considered with respect to other system parameters.

There are several techniques discussed in the open literature that avoid both the mutual coupling effects and the different implementation mismatches introduced by antenna sidelobe emission [24]-[28]. This is because it is more difficult to model the effects that the radio channel has on the real radiation pattern due to its volatility. This research focuses on these issues and adopts a solution using a simple concept of the effective radiation pattern (ERP) with Friis equation.

Friis equation shows that, the product of the transmitting antenna gain $G_t$ and the transmitter power $P_t$ is the most important result as far as the received power, $Pr$, is concerned. Any introduction of losses by the transmission line has to be added to that of the antenna system and the total losses accounted for. This is the reason why a transmission system is often represented by its effective isotropic radiated power (EIRP):

$$EIRP = P_t G_t e_{TL}, \qquad (9)$$

Where, $e_{TL}$ is the loss efficiency of the transmission line connecting the transmitter to the antenna. Usually, the EIRP is given in dB, in which case (9) becomes:

$$EIRP_{dB} = P_{tdB} + G_{tdBi} + e_{TL,dB} \qquad (10)$$

Bearing in mind that $P_{in,t} = e_{TL}P_t$ and $G_t = 4\pi U_{max,t}/P_{in,t}$, the EIRP can also be written as

$$EIRP = 4\pi U_{max,t}, \text{W} \qquad (11)$$

It is obvious from (11) that the EIRP is an imaginary amount of power that is emitted by the isotropic radiator to produce the peak power density that is seen in the direction of the maximum antenna gain. It is also clear from (9) that, the EIRP is greater than the power that is actually needed by the antenna to achieve a speculated amount of radiation intensity in its direction of maximum radiation. Hence we will model the antenna radiation power using the concept of effective radiated power (ERP).

The analytical ERP solution is developed with an assumption that the radiation pattern ($G_{real}(\varphi)$) is produced after the effect of the radio channel. Effective beamwidth ($BW_{effect}$) and effective Sidelobe suppression level ($SLL_{effect}$) are used to reflect the modified beamwidth and average sidelobe level of an ideal radiation pattern, if multipath is taken into consideration, both could be calculated via a cost function minimization that best fits the ERP with the real radiation pattern. The ERP concept assumes some factors such as the necessary number of antenna array elements, appropriate array geometry and adaptive algorithms which make up the beamforming capability that would produce radiation patterns with the desired peculiarity in relation to beamwidth and average sidelobe level. In measuring the gain of an antenna, authors in [29] started by modeling the azimuth ($\varphi$) radiation pattern ($G_{real}(\varphi)$) by spreading the ideal antenna pattern ($G_{ideal}(\varphi)$) over the environment power azimuth pattern ($A(\varphi)$).

The real radiation pattern on the antenna can also be modeled by determining the convolution of the ideal radiation pattern with the environment power azimuth pattern. Using this approach, the radiation pattern can be expressed as:

$$G_{real}(\varphi) = \oint_{\varphi} G_{ideal}(\varphi) \cdot A(\varphi - \varphi_0) \, d\varphi \qquad (12)$$

Following the approach in [30], the best way to model the power azimuth spectrum (PAS) around the BS for urban environments is to find the Laplacian distribution (LD). The LD of the environment power azimuth $A(\varphi)$ can be solved as:

Let $f(\varphi) = A(\varphi - \varphi_0)$

Then LD of $f(\varphi)$ denoted as $F(s) = \int_0^\infty f(\varphi)e^{-(\varphi-\varphi_0)\sigma} \, d\varphi$

which implies $F(s) = \int_0^\infty A(\varphi - \varphi_0) \cdot e^{-(\varphi-\varphi_0)\sigma} \, d\varphi$

$$F(s) = \frac{1}{\sqrt{2}\sigma} e^{-\sqrt{2}|\varphi-\varphi_0|}/\sigma \qquad (13)$$

Substituting $A(\varphi)$ in (12) with its LD in (13), the antenna radiation pattern can be rewritten as:

$$G_{real}(\varphi) = \frac{1}{\sqrt{2}\sigma} \oint_{\varphi} G_{ideal}(\varphi) \cdot e^{-\sqrt{2}|\varphi - \varphi_0|/\sigma} \, d\varphi \qquad (14)$$

Where, $\sigma$ is the angular spread (AS). Let us assume an ideal $N$ element linear antenna array with an element distance of $\lambda/2$ and bore-sight radiation. The ideal antenna radiation pattern ($G_{ideal}(\varphi)$) for the $N$ array antenna can be found by solving for the area of mainlobe. The ideal radiation pattern is calculated as:

$$G_{ideal}(\varphi) = \left| \frac{\sin((N/2)\pi \cos \varphi))}{(N/2)\pi \cos \varphi} \right| \qquad (15)$$

The impact of the environment azimuth power profile on $G_{ideal}(\varphi)$ can be calculated using (12)-(15) as:

$$G_{real}(\varphi) = \frac{1}{\sqrt{2}\sigma} \int_0^\pi \left| \frac{\sin((N/2)\pi \cos \varphi)}{(N/2)\pi \cos \varphi} \right| \cdot e^{-\sqrt{2}|\varphi - \varphi_0|/\sigma} d\varphi \qquad (16)$$

Where, $\varphi \in (-\pi, \pi)$

The antenna analysis in this research consider only the upper part of the antenna array pattern; that is, $\varphi \in (0, \pi)$. The aim is to use a simple step function with the sidelobe level (SLL) and the effective beamwidth (BW) to help model the antenna radiation pattern diagram with its effective radiated power(ERP) given by that step function. The ERP can only have two values, that is:

- The ERP=1, if the emission from the sidelobe of the antenna of the adjacent cell falls within the main lobe of the victim MIMO antenna.
- The ERP=$10^{-SLL/10}$, if the emission from the sidelobe of the BS of the adjacent cell falls outside the main lobe of the victim antenna.

The total amount of interference on the MIMO antenna from the sidelobe emissions depends on the number of interfering Base Station antennas having the N element MIMO antenna in their side lobe emission path. Since the antennas are uniformly distributed in the multicarrier system in Fig. 2, the probability of ERP=1 depends only on the antenna array's main lobe beamwidth, b=BW/2. Hence, the ERP values are given as:

$$G_{ERP}(\varphi) = f(x)$$

$$f(x) = \begin{cases} 1, & \varphi \in \left[\varphi_m - \dfrac{BW}{2}, \varphi_m + \dfrac{BW}{2}\right] \\ 10^{\frac{SLL}{10}} & \varphi \in \left[0, \left(\varphi_m - \dfrac{BW}{2}\right) \cup \left(\varphi_m + \dfrac{BW}{2}\right), \pi\right] \end{cases} \qquad (17)$$

Where, $\varphi_m$ is the pointing angle of the main lobe. In order to provide the best fit between $G_{real}$ and $G_{ERP}$ the BW and SLL parameters must be defined in a way that the *cosine* function given in (16) is minimized. As such:

$$\{BW_{effect}, SLL_{effect}\}$$
$$= \underset{BW \in (0,\pi), SLL \in (a,0)}{\operatorname{argmin}} \int_0^\pi |G_{ERP}(\varphi) - G_{real}(\varphi)| \, d\varphi \qquad (18)$$

Where $a$ is the minimum sidelobe level value that is utilized to outline the search area for the effective SLL. The model in (18) will enable us determine the amount of added suppression required to reduce interference within the deployment area. The results of this solution will be used to determine appropriate sidelobe emission suppression factors during the deployment stage. This suppression value will be used in the antenna simulation parameters to simplify the coverage analysis and yet produce realistic performance estimations during the radio planning phase later in this work.

### C. Received Power and Network Capacity Simulation Methodology

As part of the coverage analysis, an estimation of the Reference Signal Received Power (RSRP) was made. The information provided by RSRP is about the signal strength but rather not the quality of the signal. RSRP is the linear average of the reference signals of the downlink across the bandwidth of the channel. Some practical uses of the RSRP in LTE networks is that it is used for handover, cell selection and cell re-selection. RSRP for usable reference signals typically varies [31]. Table I gives a description of the different ranges of RSRP which is used for our coverage analysis.

TABLE I.    RSRP REFERENCE RANGE

| RSRP Power (dBm) | Description |
|---|---|
| < -90 | Excellent/Near cell |
| -90 to -105 | Good/Mid-cell |
| -106 to -110 | Fair/Cell edge |
| -110 to -120 | Poor |

TABLE II.    ANTENNA RADIATION PATTER SIMULATION PARAMETERS

| Frequency Range | 2300 MHz - 2700MHz |
|---|---|
| Angular Spread | 0, 10, 20 and 30 |
| Number of elements | 320 |
| VSWR | ≤1.5 |
| Input Impedance | 50Ω |
| Gain | 18 dBi±0.5dBi |
| Polarization | ±45° |
| Horizontal Beamwidth (3dB) | 60±5° |
| Vertical Beamwidth (3dB) | 7°±0.5° |
| Electrical Downtilt | 2° |
| Isolation Between Ports | ≥30dB |
| Front to Back Ratio | ≥30dB±2 dB |
| Cross Polarization Ratio | ≥18dB |
| Null-Fill | ≤18dB |
| Max, power | 250W |

Generally, RSRP is given in terms of Received Power (RSSI). RSSI refers to the wideband power of the user equipment (UE) that is received in total. It is made up of power from serving cells as well adjacent cell interference, thermal noise. This helps us to determine the interference from other cells. RSRP is modelled as [32]:

$$RSRP\ (dBm) = Received\ power - 10 * log\ (12 * N) \quad (19)$$

Where $N$ denotes the number of resource blocks which is dependent on the bandwidth. From the allocated 10 MHz bandwidth for the network deployment in Table II, the number of resource blocks will be 50. It implies that (19) can be rewritten as:

$$RSRP\ (dBm) = Rceived\ Power\ (dBm) - 27.78 \quad (20)$$

This implies that RSRP will be about 27dBm lower than RSSI value. Therefore RSRP coverage simulation results will be calculated as:

$$RSRP = D_r(\theta_r, \varphi_r) \cdot \frac{\lambda^2}{4\pi} \cdot \frac{P_t}{4\pi R^2} e_t D_t(\theta_t, \varphi_t) \cdot e_r |\hat{\mathbf{p}}_t \cdot \hat{\mathbf{p}}_r|^2 - 27.78 \quad (21)$$

In this work Genex Unet will be used to predict RSRP distribution using (21) and urban parameters.

In evaluating the capacity of LTE network, the fluid model proposed by authors in [33] was used. The signal interference plus noise ratio for PUSc scheme is calculated as [33]:

$$YPUSC(r) = \frac{K\sqrt{3}}{\pi}(\eta - 2)(2\sqrt{K} - r)^{-2}(\frac{2\sqrt{K}}{r-1} - 1)^\eta \quad (22)$$

Where

$K = Reuse\ factor$

$\eta = interference\ to\ nise\ ratio$

$r = cell\ radius$

Hence the capacity for the PUSC 1x3x3 scheme can be evaluated as [33]:

$$C_{PUSc}(k) = \log_2[1 + YPUSC(r)] \quad (23)$$

The capacity estimation will be done with (23) under varying reuse schemes and discussed in subsequent sections.

## VI. SIMULATION PARAMETERS AND FIELD MEASUREMENT SETUP

The radiation pattern of the MIMO antenna is simulated using the parameters in Table III. Unsuppressed sidelobe emissions distort the performance of MIMO antenna systems used in deploying multicarrier networks. In order to accurately evaluate the performance of any MIMO antenna configurations used in a realistic network deployment scenario, this effect must be modeled and simulated.

The received power and throughput of three MIMO configurations used in a practical network deployment scenario are also simulated using the adopted step function, coverage and capacity estimation models with the parameters in Table II.

TABLE III.     RADIO NETWORK PLAN SIMULATION PARAMETERS

| Parameter | Description |
|---|---|
| Carrier Frequency | 2620-2630 MHz |
| System Bandwidth | 10MHz |
| OFDM Symbol time | 102.8 ms |
| Transmit Power | 43dBm |
| Receiver Sensitivity | -110 dBm |
| Mobile Receiver Height | 0.5m |
| Base Station antenna height | 35 m |
| Transmitter antenna gain | 18.5 dBi |
| Area | Urban |
| LTE Duplex mode | TDD |
| MIMO Scheme | 2X2, 4X4, 8X8 MIMO |
| Downlink multiple access | OFDMA |
| Azimuth (degree) | 0/120/240 |

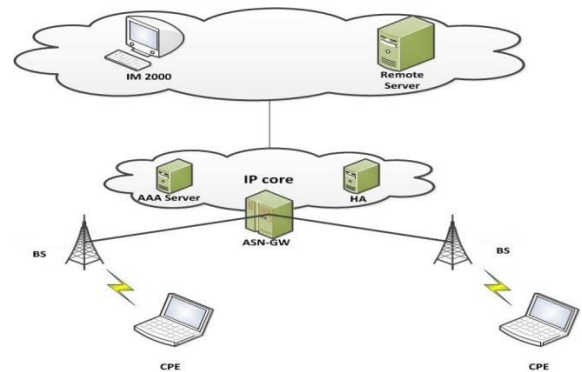

Fig. 4.   Field experimental setup.

The field experimental trial results will be presented after the simulation results have been discussed. The objective of the field experimental test was to compare simulated and actual deployment results. The field measurement setup is shown in Fig. 4. The measurement set up is comprised of a GPS connected to an LTE customer premise equipment (CPE) with Genex-probe software installed on it, and a Kathrein omnidirectional with 2x2 MIMO configuration.

The data was collected by undertaking a drive test and querying a remote server using IM2000, a live network monitoring tool. The data was collected at hourly intervals in the month of December when traffic on networks in Ghana are generally high. The field measurements were done for only 2x2 MIMO because the deployed LTE network under study uses adaptive 2x2 MIMO configuration.

## VII. SIMULATION RESULTS AND ANALYSIS

This section presents the simulation results obtained for network coverage and throughput when the theoretical coverage, sidelobe power and coverage models are implemented in NEC and Genex Unet.

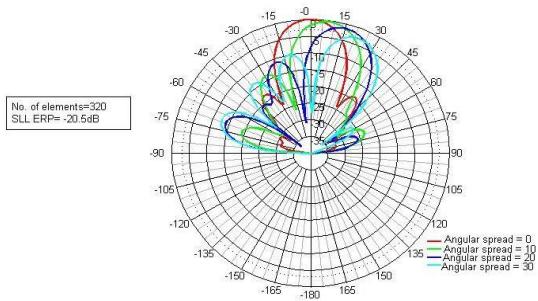

Fig. 5.   Antenna radiation pattern simulation result.

### A. Antenna Radiation Pattern

In this section, the step function in (18) is used for evaluating the effective beamwidth and the suppression factors of the antenna radiation using NEC simulation tool. Following the conclusions derived in [34], the Angular Spread (AS) for a bad urban microcell radio node was chosen as 330. Therefore, we employed in the Laplacian distribution different AS values of the MIMO antenna and perform the analysis to model the real radiation patterns shown in Fig. 5 using the simulation parameters in Table III.

Results of the vertical radiation patterns shown in Fig. 5 indicates that the limiting case for the minimum value of side lobe level is when the angular spread is under the selected antenna element of 320. From results, when the angular spread is almost 0 and the number of linear array elements is 320, the corresponding effective radiated sidelobe power level is $-20.5$ dB. From this result, a value of optimal sidelobe level suppression of $a \geq -20.5$ dB is recommended for an antenna with $n \geq 320$ for network deployment under bad urban conditions. As a result, a value of $a = -20$ dB and $n=320$ was used to model practical deployment scenarios presented in this research.

### B. RSRP Simulation Results

To find out the maximum cell range under the deployment scenario, a single site simulation was conducted for 2X2 MIMO, 4X4 MIMO and 8X8 MIMO. Simulation was performed for users in urban areas within the Accra metropolis operating at 2600MHz.

The RSRP levels were randomly distributed ranging between -50 to -140dBm for the three MIMO configurations. The receiver sensitivity required for minimum performance was obtained at -110 dBm which implies that a cell usable range of 2km was achievable using 2X2 MIMO as indicated in Fig. 6. The minimum threshold sensitivity however was -133 dBm which implies that the network can be sensed at about 3 km using 2X2 MIMO but a user will not enjoy expected throughput after 2km away from the cell site.

For the coverage simulation of a single 4x4 MIMO configuration shown in Fig. 7, RSRP levels were also randomly distributed ranging between -140 to -50 dBm similar to that of the 2x2 MIMO configuration. The receiver sensitivity required for minimum throughput was obtained at -110 dBm which implies that a cell usable range of 2.5km is achievable using 4X4 MIMO under the deployment scenario.

The detectable cell range for the adaptive 4x4 MIMO configuration was obtained at 3.5km.

The receiver sensitivity required for maximum throughput is -110 dBm which implies that the cell's usable and detectable ranges of 2.8km and 4km respectively were achievable using 8X8 MIMO as summarized in Fig. 8. Cell usable ranges of 2km, 2.5km and 2.8km were obtained for 2X2, 4X4 and 8X8 MIMO respectively. It can be observed that as the number of transmitting antennas increases, the coverage of cell is enhanced as shown in the comparison results in Fig. 9.

The main purpose of the initial planning process was to investigate and develop efficient and low cost radio access systems to provide users in an urban environment a full range of broadband services. The final radio plan in Fig. 6 for the deployed 2X2 MIMO configuration helps achieve this objective through efficient utilization of radio frequency bands and optimization of transmission capacities for the different variety of users within the pilot network. To predict the radio wave propagation in the network, the planning tool took into account the antenna radiation patterns results. Since the information used to determine the simulation parameters were obtained from the network planning engineers, the result in Fig. 6 produces similarly structured cell plans as the one being used by the network operator.

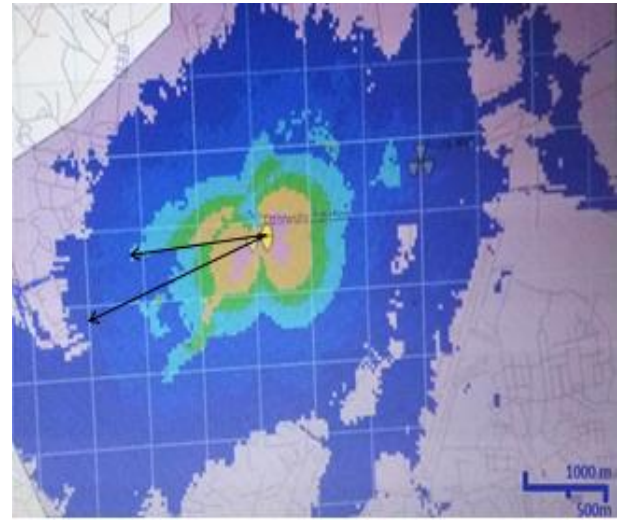

| Color | Legend | Percentage |
|---|---|---|
| | [-50,0] | 0.00 |
| | [-70,-50] | 0.01 |
| | [-80,-70] | 0.75 |
| | [-90,-80] | 3.38 |
| | [-105,-90] | 4.12 |
| | [-120,-105] | 6.91 |
| | [-120,-110] | 21.39 |
| | [-140,-120] | 63.43 |

Fig. 6.   RSRP coverage performance of a single site that uses 2X2 MIMO (dBm).

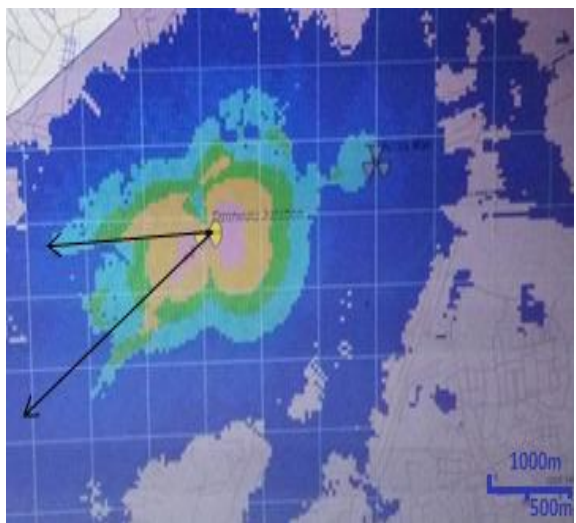

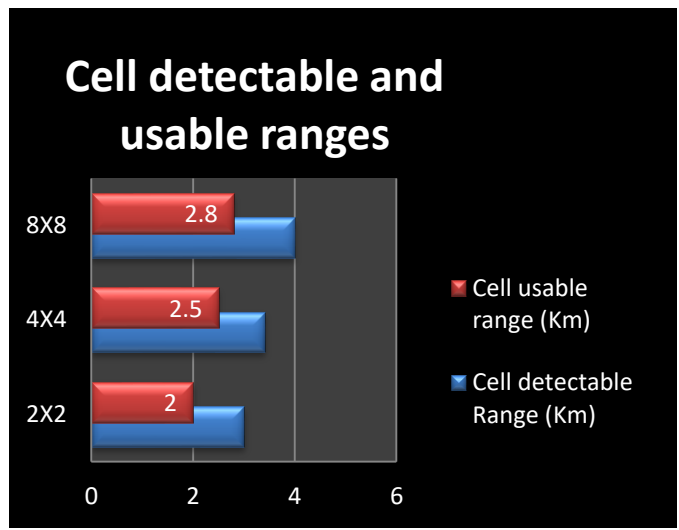

Fig. 9. Graph showing the cell ranges for single site for 2X2, 4X4 and 8X8 MIMO.

| Color | Legend | Percentage |
|---|---|---|
| | [-50,0 ] | 0.00 |
| | [-70,-50] | 0.06 |
| | [-80,-70] | 1.24 |
| | [-90,-80] | 4.43 |
| | [-105,-90] | 4.96 |
| | [-110,-105] | 8.58 |
| | [-120,-110] | 20.56 |
| | [-140,-120] | 60.07 |

Fig. 7. RSRP coverage performance of a single site using 4X4 MIMO (dBm).

The simulated radio plans in this research however achieves a coverage level above 97% as compared to the 90-92% which was obtained by the network engineers during the planning stage. The increase in the coverage level could be attributed to the usage of sidelobe suppression in the final radio network plan. This in turn implies that the modeling of antenna sidelobe suppression factors in any multicarrier network system will maximize the overall network service level of its MIMO configuration.

*C. Throughput Simulation*

From the network simulation parameters in Table II, at 10MHz, the OFDMA symbol time is 102.8microseconds and so there are 48.6 symbols in a 5 milliseconds frame. Of these, 1.6 symbols are used for TTG (Transmit to Transmit Gap) and RTG (Receive to Transmit Gap) leaving 47 symbols. If n of these symbols are used for DL, then $47 - n$ are available for uplink. Since DL slots occupy 2 symbols and UL slots occupy 3 symbols, it is best to divide these 47 symbols into a Time Division Duplexing (TDD) Downlink/Uplink (DL/UL) split ratio such that $47 - n$ is a multiple of 3 and n is of the form $2k + 1$. Based on the propagation mode, the required frequency reuse scheme and the DL/UL channel ratio are used. For a capacity limited network, a 35:12 TDD split ratio is used to serve the users in the network. When the sole purpose of the network is to cover a larger area, a DL/UL ratio of 26:21 is used. Under the adaptive 2x2 MIMO deployment scenario, the maximum average DL and UL throughput per sector are 28.51 Mbps and 11.98 Mbps respectively for a coverage limited network using a 26:21 DL/UL ratio and a PUSC 1x3x3 reuse scheme. The results of the LTE capacity simulation results in this work is compared with the results presented in [15] for the first successfully deployed WiMAX network in Ghana in Fig. 10. It can be seen that, the results of the LTE throughput simulation indicates an impressive performance of a Wireless network never seen before in the sub-region.

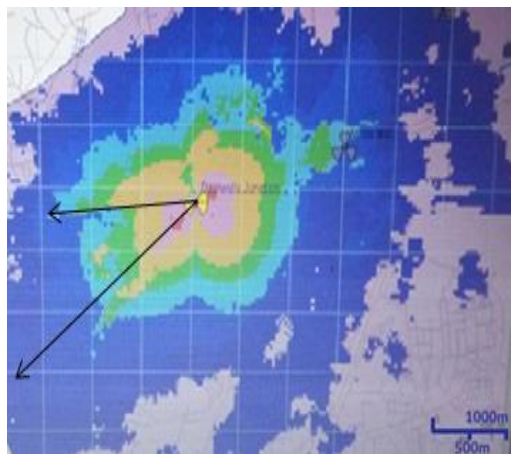

| Color | Legend | Percentage |
|---|---|---|
| | [-50,0 ] | 0.00 |
| | [-70,-50] | 0.14 |
| | [-80,-70] | 1.66 |
| | [-90,-80] | 5.77 |
| | [-105,-90] | 6.15 |
| | [-110,-105] | 8.72 |
| | [-120,-110] | 20.23 |
| | [-140,-120] | 57.33 |

Fig. 8. RSRP coverage performance of a single site using 8X8 MIMO (dBm).

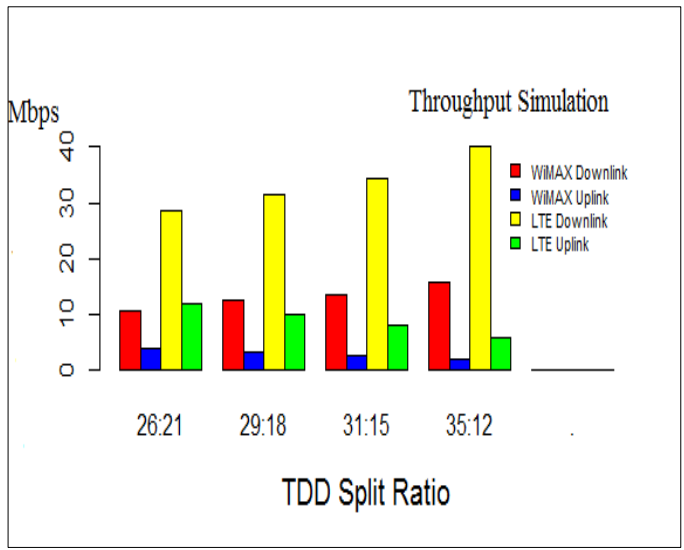

Fig. 10. Throughput simulation under varied TDD Split ratios.

Due to this impressive performance of the LTE network, the DL/UL ratio can also be chosen to be flexible to adapt to the differing demands of wireless services, whether to support high bandwidth, low latency, bursty traffic, ultra-reliable services based on the type of application being deployed in the network. For example, in deploying a mobile Telemedicine network which requires a higher UL throughput, a DL/UL split ratio of 26:21 is best suited for such network deployments whiles a DL/UL ratio of 36:12 can be used to deploy a network that provides online streaming services. The simulation performance of the deployed LTE network gives the Ghanaian mobile data subscriber hope of optimizing and leveraging the use of communication resources to support diverse business models, interactive applications and solve the limited wireless connectivity problems widely experienced in Ghana.

This research further performed average throughput per sector simulation for the pilot area under varied MIMO configurations. This enabled comparison of the results of the deployed 2X2 MIMO configuration with alternative 4X4 and 8X8 MIMO configuration. This simulation was carried out using 1000 users for each cell site. 20 cell sites were put together in this simulation with each cell site having three sectors to be able to fairly represent the number of cell sites serving end users within the pilot network. Fig. 11 presents 2x2 MIMO throughput distribution simulation. The area as indicated by the red ink has between 0-1 Mbps throughput coverage using 2X2 MIMO. Very poor power levels of -120 to -110 dBm were observed in this area. Throughput distribution however varied between 0 and 40 Mbps in the simulation. Average downlink throughput of 8.324 Mbps was obtained using 2X2 MIMO.

The throughput simulation results for adaptive 4x4 MIMO antenna configuration is summarized in Fig. 12. Using 4X4 MIMO, the areas indicated by the highlighted regions on Fig. 12 show very similar throughput results to 2X2 MIMO within those areas with coverage gaps. However, throughput values of 0-1 Mbps that were observed in 2X2 MIMO was considerably reduced in 4X4 MIMO. On close observation 4X4 MIMO gave better average throughput results over 2X2 MIMO. Using 4X4 MIMO an average downlink throughput of 9.1782 Mbps was obtained.

Fig. 13 also presents 8X8 MIMO throughput distribution simulation for the study area. Form the results, it was realized that an average downlink throughput of 9.517 Mbps was obtained using 8X8 MIMO. 8X8 MIMO produced the best throughput coverage for high throughput values of 10 - 30 Mbps. Throughput values obtained from all simulation gives a correlation to RSRP coverage results. Areas with low RSRP - 120 to -110dBm had low throughput 1-5Mbps values. Areas with high RSRP coverage levels -100 to -50dBm had high throughput 5-40 Mbps. Generally the throughput behavior varied randomly. The average values however gave us an idea about the throughput performance using different MIMO schemes and this comparison as has been shown in Fig. 14. The downlink throughput prediction for the cluster analysis generally gave a higher average throughput for 8X8 MIMO. 2X2 MIMO gave the least average throughput per sector. 8X8 MIMO with the highest number of transmitting antennas gave the highest average throughput compared to 2X2 and 4X4 MIMO. The percentage coverage for users under the 2x2 MIMO simulation scenario was better than that of the adaptive 4x4 MIMO configuration with 2X2 MIMO achieving 60.41% of area having throughput values between 1 - 40Mbps as against 55.87% achieved by the 4x4 MIMO configuration. This result seems to support the conclusions by authors in [13] that 2x2 MIMO seems to perform better in the peculiar Sub-Saharan terrain profile than adaptive 4x4 MIMO configuration.

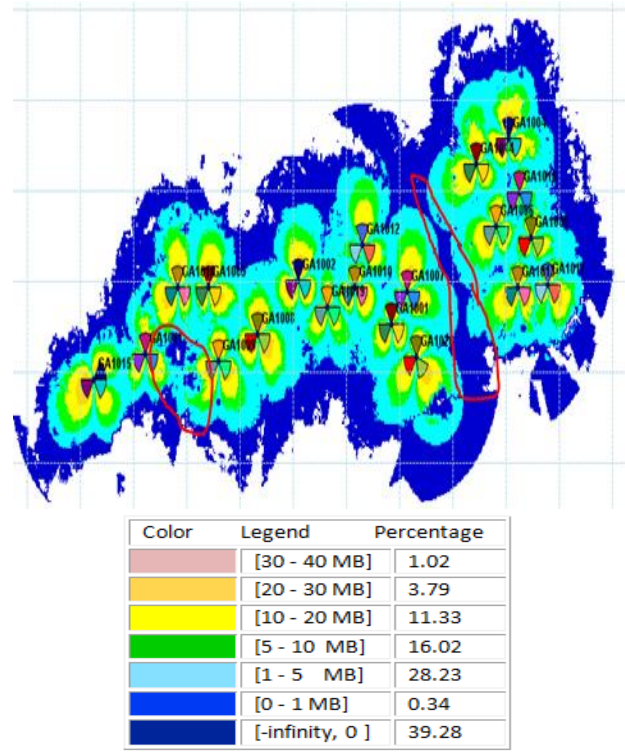

| Color | Legend | Percentage |
|---|---|---|
| | [30 - 40 MB] | 1.02 |
| | [20 - 30 MB] | 3.79 |
| | [10 - 20 MB] | 11.33 |
| | [5 - 10  MB] | 16.02 |
| | [1 - 5   MB] | 28.23 |
| | [0 - 1 MB] | 0.34 |
| | [-infinity, 0 ] | 39.28 |

Fig. 11. Throughput distributions simulation for 2X2 MIMO.

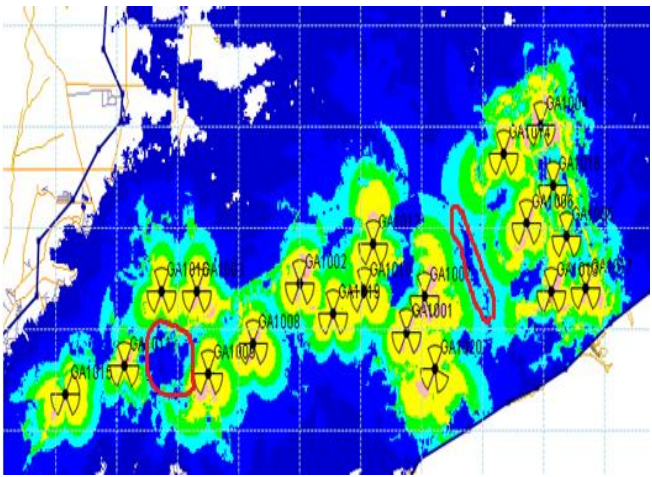

| Color | Legend | Percentage |
|-------|--------|-----------|
| | [30 - 40 MB] | 0.87 |
| | [20 - 30 MB] | 3.98 |
| | [10 - 20 MB] | 12.36 |
| | [5 - 10  MB] | 18.25 |
| | [1 - 5  MB] | 20.41 |
| | [0 - 1 MB] | 0.07 |
| | [-infinity, 0 ] | 44.07 |

Fig. 12. Throughput distributions simulation for 4X4 MIMO.

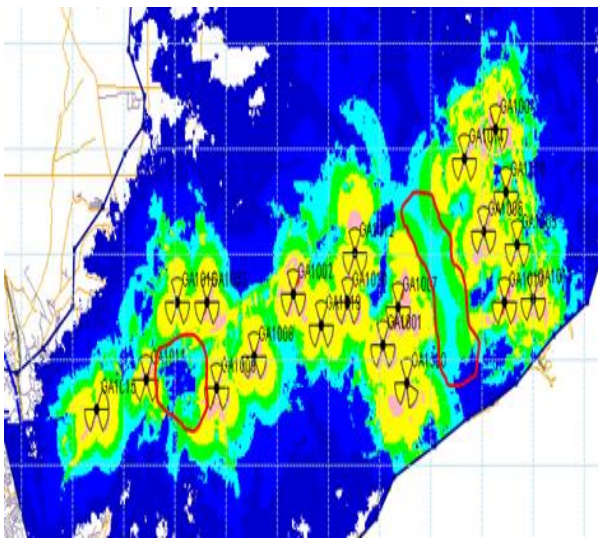

| Color | Legend | Percentage |
|-------|--------|-----------|
| | [30 - 40 MB] | 0.87 |
| | [20 - 30 MB] | 4.11 |
| | [10 - 20 MB] | 12.60 |
| | [5 - 10  MB] | 18.24 |
| | [1 - 5  MB] | 24.80 |
| | [0 - 1 MB] | 0.1 |
| | [-infinity, 0 ] | 39.28 |

Fig. 13. Throughput distributions simulation for 8X8 MIMO.

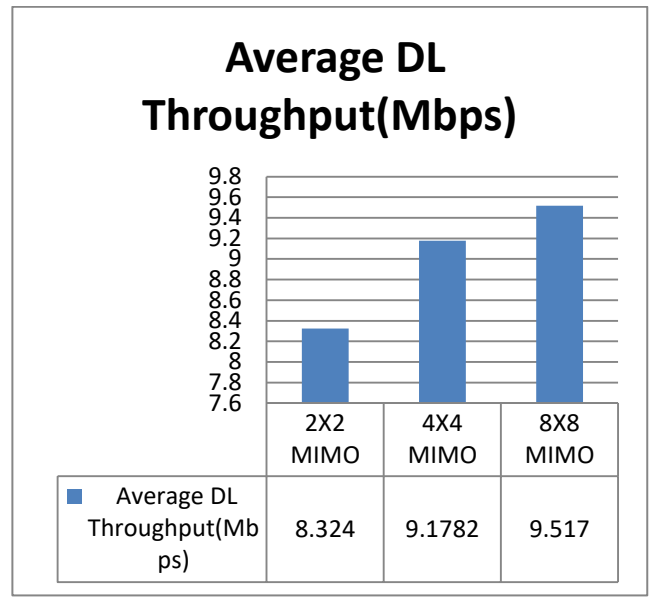

| | 2X2 MIMO | 4X4 MIMO | 8X8 MIMO |
|---|---|---|---|
| Average DL Throughput(Mbps) | 8.324 | 9.1782 | 9.517 |

Fig. 14. Average throughput comparison for 2X2, 4X4 and 8X8 MIMO.

## VIII. FIELD MEASUREMENT TRIALS RESULTS

This section presents the results collected during the field experimental trials. A comparison analysis has subsequently been done.

### A. RSRP Measurements Results

Field results was collected for 2X2 MIMO 4G LTE. Fig. 15 presented below is a graphical representation of the drive test results on the reference signal received power conducted at the test route.

RSRP distribution differs at different points within the coverage area. Very good signal levels up to -110 dBm were measured up to about 2.5 km on route 2. Poor power levels below -110 dBm were not experienced on route 2.

However along route 1, poor levels lower than -110 dBm were experienced at about 1.5 - 2 km. Drive test was conducted on Route 2 which is on the N1 motorway. Here signals traversed free space along the motorway and could go as far as 2.5 km as shown by Fig. 9 on the ash route. Route 1 however finds itself on streets surrounded by cluster of buildings as shown by Fig. 16 and good signal level up to -110 dBm was received as far as 1.4 km. High rise buildings along the street obstructed the signals and increased the path loss reducing the signal received power.

From the drive test conducted in this sector, network coverage can reach a range of about 2.5 km with no obstructions but limited coverage of 1.4km in the presence of obstructions.

Simulation and field collected results for Route 1 showed correlation as shown in Fig. 10. In both cases signals deteriorated after 1.4 km. Simulation results were better in terms of RSRP level than field collected results. The buildings served as obstruction for the signal on this route and impeded the cell range.

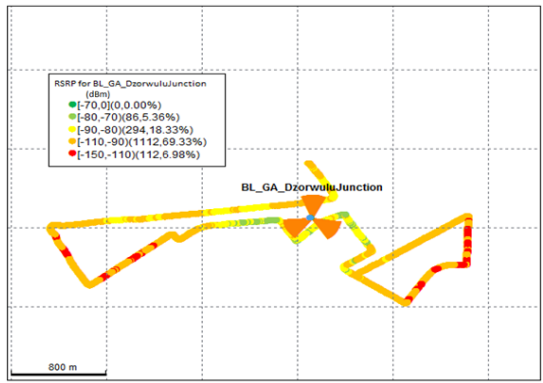

Fig. 15. RSRP distribution at test route (dBm).

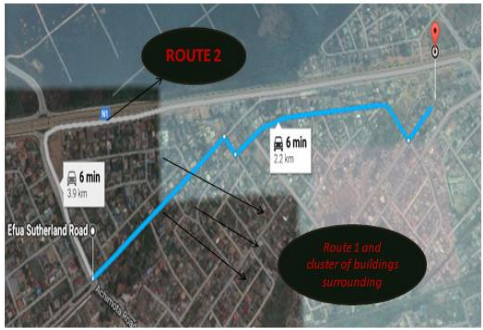

Fig. 16. Overview of selected drive test routes.

Simulation results on Route 2 proved that RSRP received power decreases with increasing distance as shown in Fig. 17. Field collected data gave similar results until after 1.5 km when the received level remained nearly constant at (-100 dBm) from this point to about 2.5 km. This shows that RF signals traverse farther distance in a free space.

From the system level simulation results, a peak downlink and uplink throughput values of up to 75Mbps and 35Mbps were obtained respectively. Field measurement using IM2000 Monitor however showed 62.318 Mbps and 10.23 Mbps for downlink and uplink respectively. The deviation of the measured peak throughput values from the expected theoretical throughput could mainly be attributed to losses at the antenna front end and the use of pathloss models which have not been corrected to suite the peculiar Sub-Saharan African terrain among other factors.

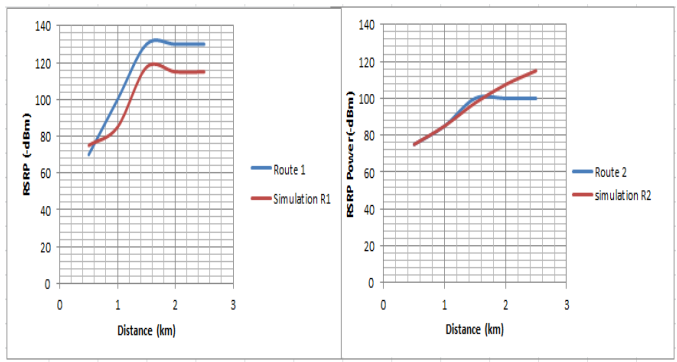

Fig. 17. Simulation against field measured coverage.

### B. Throughput Field Collected Results

The results of the peak downlink and uplink measurements are summarized in Fig. 18 and 19.

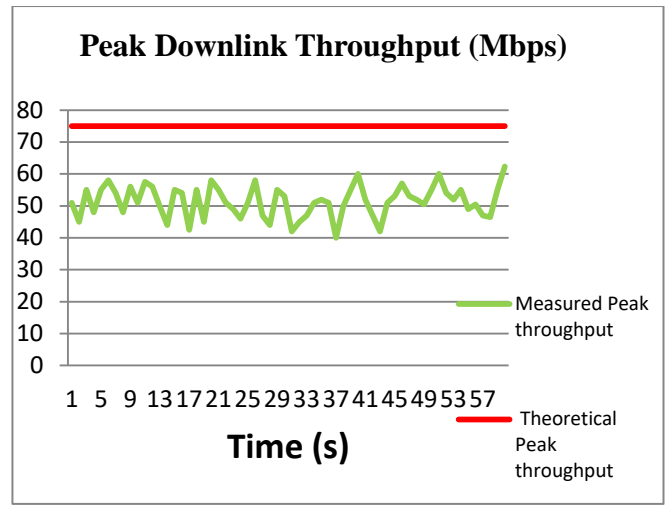

Fig. 18. Graph showing peak downlink throughput from drive test.

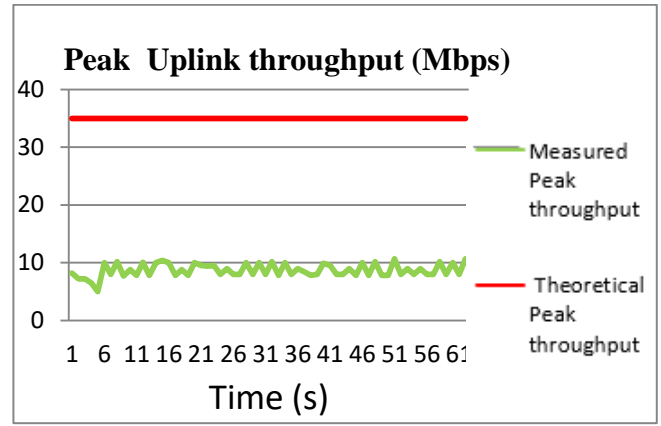

Fig. 19. Graph showing peak uplink throughputs from drive test.

Average throughput per sector measurements was also carried. The summarized results of the average throughput per sector recorded in this study period are presented in Fig. 20. Simulation results gave maximum per sector throughput values up to 40Mbps under a 35:12 TDD split ratio. Results obtained from both field data collection and simulation showed variation in throughput. The maximum simulated was 40 Mbps whiles field experimental results showed 29.9 Mbps. During the field measurement, it was observed that the throughput values varied depending on the traffic load, number of concurrent users and the distance of the user away from the cell site.

The maximum measured average downlink throughput per sector of 29.9 Mbps showed better performance of the deployed 4G technology when compared to the 9.52 Mbps obtained in the performance evaluation of WiMAX networks in the same study area which was reported in [15]. The comparison of LTE results and the results obtained in the same area for WiMAX is presented in Fig. 21.

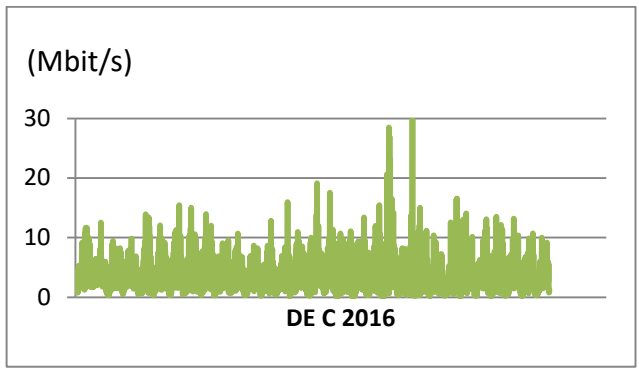

Fig. 20. Downlink average throughput per sector for the measurement period.

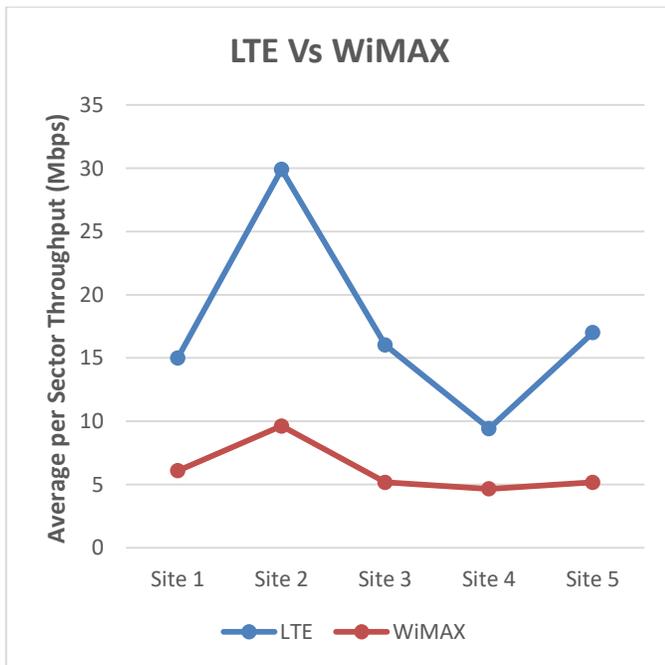

Fig. 21. Measured LTE throughput performance against documented performance of WiMAX in the survey area.

The performance of the two 4G technologies in the pilot area was observed to be directly related to the user densities in the survey area. Site 4 had the highest recorded number of subscription at 15,000 users during the measurement period with Site 2 having the least number of users at 5,113. Even though there is limited documented results on the performance evaluation of wireless networks in the sub-region, the results obtained from the experimental measurement in this work confirm LTE as superior wireless broadband technology which is be capable of supporting last mile broadband solutions in the sub-region.

## IX. CONCLUSION

This paper has presented throughput and coverage analysis using various MIMO antenna configuration schemes. Simulations showed 8X8 MIMO to perform better over 2X2MIMO and 4X4 MIMO in terms of throughput and coverage. However, it was seen that, the general performance of adaptive 2x2 MIMO was better than 4x4 MIMO configuration. Field data collected on deployed 2X2 MIMO 4G LTE operating in the 2600 MHz band showed a measured maximum average throughput per sector of 29.9 Mbps and peak downlink throughput of 62.318 Mbps for users within a cell range of 2.5km away from the Base Station. Based on the results in this work, it can be concluded that, 4G LTE therefore is capable of providing the ever increasing broadband demand of Ghanaians when comparison is made with the throughput requirement needed to support data-centric broadband applications.

Future works on antenna configuration could concentrate on studying the effect of the use of propagation pathloss models which don't have offset parameters specified for the peculiar Sub-Saharan African terrain on the coverage gaps found in the detectable cell range.